# Recent topics on metastability, hysteresis, avalanches, and acoustic emission associated to martensitic transitions in functional materials.

Martin-Luc Rosinberg [1] and Eduard Vives [2,3]

(1) *Laboratoire de Physique Théorique de la Matière Condensée,*
*Université Pierre et Marie Curie,*
*4 Place Jussieu, 75252 Paris,*
*France*
(mlr@lptl.jussieu.fr)

(2) *Department of Physics,*
*University of Warwick,*
*Coventry CV4 7AL,*
*United Kingdom*

(3) *Departament d'Estructura i Constituents de la Materia,*
*Universitat de Barcelona,*
*Facultat de Física, Martí i Franquès 1, 08028 Barcelona,*
*Catalonia*
(eduard@ecm.ub.es)

I INTRODUCTION

Functional materials are based on the interplay of different ferroic properties like ferroelasticity, ferromagnetism, ferroelectricity, etc. In order to obtain a sufficiently large response to the external excitation, these materials are typically tuned so to cross a first-order phase transition in which one or several order parameters (strain, magnetization, polarization, etc…) exhibit a macroscopic discontinuity. It is thus important for applications to understand the dynamics of first-order phase transitions (FOPT) in such materials.

FOPT in solids hardly occur at thermal equilibrium. Typically, the energy barriers involved in the transition are very large compared to thermal fluctuations so that the order parameters evolve following metastable trajectories. The transitions are then called athermal, and instead of being sharp like in ideal first-order phase transitions, they extend over a broad range of the driving parameter and show hysteresis. In many cases the hysteresis (or at least a part of it) cannot be suppressed by driving the system more slowly because it is not related to the fact that the system cannot respond instantaneously due to relaxational delay. This kind of hysteresis is usually called rate-independent hysteresis.

The high energy barriers have two origins: on the one hand, real materials always exhibit some amount of quenched disorder that determines the nucleation sites and can thus strongly affect the metastable path. On the other hand, when one of the order parameters involved in the transition is strain (like in martensitic transformations), a complex microstructure naturally arises at the FOPT due to the symmetry differences between the parent and product phases. This also strongly affects the metastable trajectory and the hysteresis.

In section II we shall introduce very simple models of athermal evolution in driven ferromagnets. They give us a global picture of the relationship between metastability and hysteresis and show that athermal FOPT in the presence of disorder proceed via avalanches. This means that the response of the system to a smooth driving consists in a sequence of discontinuous jumps of the order parameter separated by periods of inactivity. Microscopically, avalanches are associated with the motion of an interface and/or with the nucleation of a domain of the new phase. These models also describe how the statistical distribution of the avalanche sizes changes with the amount of disorder, how avalanches and hysteresis depend on the driving mechanism, temperature, driving rate, number of cycles through the transition, etc. In magnetic materials, the signature of the avalanche dynamics is the so-called Barkhausen noise which can be monitored by using a pick-up coil. Similar phenomenology is observed in other disordered systems, for instance in ferroelectrics or superconductors.

In section III we will focus on the ferroelastic case of structural phase transitions where avalanches can be recorded as acoustic emission (AE) events. These are produced when an interface separating two different crystallographic structures advances producing an elastic wave (typically with frequencies in the ultrasonic range) which propagates through the material and can be recorded at the surface by an appropriate transducer. Although a satisfactory theoretical framework to interpret the results of AE experiments is still lacking, the interaction between experiments and theory over the past 15 years has been intense and fruitful. We hope that this short review will help to clarify the status of some recent advances and contribute to further progress.

II <u>WHAT CAN WE LEARN FROM SIMPLE MODELS?</u>

On general grounds, one can relate the intermittent and hysteretic response of a disordered system to slowly changing external conditions to the existence of a corrugated (free) energy landscape. At low enough temperature this landscape is indeed characterized by an enormous number of local minima (or metastable states) and the energy barriers are so large that thermally activated processes play a negligible role. As stressed above, true equilibrium is then never reached on experimental time scales and the system can only move from one metastable state to another as the external control parameter (e.g. strain, magnetic field, pressure, or temperature) is changed and the initial state in which the system was trapped becomes unstable. This collective event (avalanche) is usually very fast, at least compared to the rate of variation of the external parameter, and this results in a jump discontinuity in the non-equilibrium response. One then often considers the so-called adiabatic limit in which the rate is merely taken to zero.

Can we go beyond such general statements? For instance, what can be said about the number of metastable states, their energy, or their magnetization (in the remainder of this section we shall most often refer to magnetic systems as illustration)? What is the relationship between the organization of the states and the shape of the saturation hysteresis loop obtained by cycling the field between large negative and positive values? What is the influence of the driving mechanism on the dynamical response (although one often controls an intensive external force or field, in other situations, e.g. in experiments with ferroelastic materials, one usually controls the strain, which is an extensive quantity, instead of the stress). What is the statistical distribution of the size and duration of the avalanches? Why is a power-law (scale-free) behaviour extending over several decades so often observed? Is this associated to a non-equilibrium critical point and then what is the range of the critical regime? In which cases do microscopic details affect large scale events and in which cases are they irrelevant?

Such questions clearly touch fundamental issues in the theory of disordered systems and to answer them (or at least some of them), it has proven useful to consider models that are simple enough to allow for a partial analytical description and for extensive numerical studies. Perhaps the simplest (and yet not fully understood) prototype is the zero-temperature nonequilibrium random field Ising model (RFIM) proposed in 1993 by J. P. Sethna, J. A. Krumshansl, and their collaborators as a model for hysteresis and crackling noise in disorder-driven first-order phase transformations [1]. This model, which contains the most important physical ingredients (quenched-in disorder, interactions, external control parameter), has been intensively studied over the past fifteen years and has been applied to many different physical situations, from fluid invasion inside porous solids to group decision making (we refer the reader to [2] for a comprehensive review). In particular, the RFIM appears to be the convenient theoretical framework to understand the hysteresis behaviour associated to the capillary condensation of gases in amorphous porous solids. In this case, the driving force is the gas pressure in the external reservoir, the order parameter is the amount of adsorbed gas inside the solid, and temperature is just a parameter that changes the topology of the free-energy landscape (but is still too low for making activated processes efficient and inducing rate-dependent hysteresis effects) [3]. A nice example coming from low-temperature physics is the condensation of helium in silica aerogels where a description in terms of random (but correlated) fields gives a rationale to the changes in the shape of the hysteresis loop with porosity and temperature [4, 5].

The RFIM is defined by the Hamiltonian

$$H = -J \sum_{<i,j>} s_i s_j - \sum_i (H + h_i) s_i \quad ,$$

(1)

where $\{s_i\}$ are N Ising spins placed on the sites of a lattice (e.g. a cubic lattice), $J > 0$ is a ferromagnetic coupling between nearest-neighbour spins, $H$ is the external field, and $\{h_i\}$ is a set of uncorrelated random fields usually drawn from a Gaussian distribution probability with zero mean and standard deviation $\Delta$.

The zero-temperature metastable evolution induced by the external field consists in a single-spin-flip dynamics: metastable states are thus defined by the condition

$$s_i = sign(f_i) \quad ,$$

(2)

where $f_i = J \sum_{j/i} s_j + H + h_i$ is the effective local field, and a spin flips when its local field changes sign. To stay in the adiabatic limit the external field $H$ is kept constant during the propagation of an avalanche. (Note that a two-spin-flip dynamics has also been considered recently to test the robustness of the model behaviour with respect to an additional relaxation process [6].) The most salient feature of the model is the existence of two regimes of avalanches depending on the amount of disorder (i.e. the value of $\Delta$). In strong disorder, spins mostly flip individually so that avalanches are of microscopic size and the magnetization curve is smooth macroscopically. On the other hand, in the low disorder regime, spins tend to flip collectively which results in a system spanning avalanche seen as a macroscopic jump in the magnetization curve. In between, there is a critical disorder $\Delta_c$ and a critical field $H_c$ at which avalanches of all sizes are observed. The avalanche size distribution then follows on long length scales a power-law behaviour as $p(S) \propto S^{-(\tau+\sigma\beta\delta)}dS$ where $\tau$, $\sigma$, $\beta$, and $\delta$ are universal critical exponents [2].

We now focus on two issues that were not discussed in [2].

1) <u>Relationship between hysteresis and the distribution of metastable states:</u>

Let us first discuss the issue of the number and distribution of metastable states in the field-magnetization plane. Since we are interested in the relationship to the hysteresis loop induced by a variation of the magnetic field H, the relevant quantity is not the total number of metastable states but the number of states with a given magnetization at a given external field $N(m,H)$ (determining the *full* topology of the energy landscape is also a very interesting and challenging issue that has received some attention in recent years [7]). On general grounds, one expects $N(m,H)$ to scale exponentially with the system size, say the number N of elementary domains, spins, etc. It is thus the logarithm of $N(m,H)$ which is an extensive quantity like the free energy. Since this quantity is sample-dependent, the physically relevant quantity is $\overline{\log N(m,H)}$, where the average, denoted here by the overbar, is taken over a representative set of disorder realizations. This leads to define the magnetization-dependent *quenched complexity* as

$$\Sigma_Q(m,H) = \lim_{N \to \infty} \frac{1}{N} \overline{\log N(m,H)} \quad .$$

(3)

Note that we consider the average of the logarithm and not the logarithm of the average (the so-called *annealed* average) because the hysteresis loop is a self-averaging quantity (sample-to sample fluctuations vanish in the thermodynamic limit) and we thus need to describe the behaviour of a *typical* sample. Considering the annealed average is misleading since there exist a certain number of atypical samples that give a finite contribution to $\log \overline{N(m,H)}$ [8,9] (on the other hand, computing the quenched complexity is much more difficult task).

The crucial point is that the hysteresis loop in the strong disorder regime (i.e. when the loop is smooth) is just the convex envelope of the set of metastable states in the field-magnetization plane and identifies with the contour $\Sigma_Q(m,H) = 0$. This is still a kind of conjectural statement but it is strongly supported by a) an exact theorem and b) analytical and numerical calculations. The exact theorem is the so-called *no-passing* rule [1] which applies to systems with ferromagnetic interactions only (or to elastic media with a convex elastic potential). The no-passing rule states that the $T = 0$ metastable dynamics conserves the partial ordering of the states: in other words, a spin never flips back when the field is varied monotonically. This is sufficient to prove the remarkable property of return point memory [1] which is observed in many experimental systems, and this also implies that there are no metastable states outside the hysteresis loop in a given disorder sample (and therefore, on average, the density of metastable states scales to zero exponentially outside the loop [8]). On the other hand, the no-passing rule does not imply that the number of metastable states is exponentially large (and therefore comparable to the total number of states) everywhere *inside* the loop. Of course, it is known experimentally that there are many metastable states inside the loop, as illustrated by the so-called `scanning' curves obtained by reversing the field (or the stress) before reaching saturation (or complete phase transformation). However, these states which are reachable by a field history starting from one of the two saturation states only represent a negligible subset of the whole set of metastable states, albeit probably the most interesting one. In fact, very little is known about their actual number [10]. Therefore, in principle, a region could exist in the vicinity of the hysteresis loop where the number of metastable states is only sub-exponential (and thus $\Sigma_Q(m,H) = 0$). Such a region, however, does not exist in the one-dimensional RFIM for which the quenched complexity has been computed analytically [8]. There is also good numerical evidence that the same is true in three dimensions [9] and we believe that this is a general feature.

The situation in the low disorder regime is different as one does not expect the envelope of the metastable states to be convex anymore. This conjecture is again based on numerical and analytical calculations of the complexity [9, 11] but it also quite naturally explains the presence of a finite jump in the magnetization curve because $m$ must always be a monotonic function of $H$ (on general grounds, the `susceptibility' $dm/dH$ must be a positive quantity (see also [12])).

These predictions are nicely illustrated by the soft-spin version of the RFIM in the infinite range limit where each spin is now a continuous variable taking values between $-\infty$ and $+\infty$ and is coupled to all other spins with coupling $J/N$, as described by the Hamiltonian

$$H = -\frac{J}{2N} \sum_{i \neq j} s_i s_j - \sum_i (H + h_i) s_i + \sum_i V(s_i)$$

(4)

where $V(s) = (k/2)[s - sign(s)]^2$ is a double-well potential that mimics the two states of the hard-spin model. The metastable states are now solutions of the equation

$$s_i - sign(s_i) = \frac{Jm + H + h_i}{k} \quad .$$

(5)

The mean-field character of the model allows one to compute the hysteresis loop and the complexity $\Sigma(m,H)$ analytically [13]. Some typical results are shown in Figures 1 and 2 (note that the magnetization does not saturate when $H \to \pm\infty$ because the spins are unbounded variables).

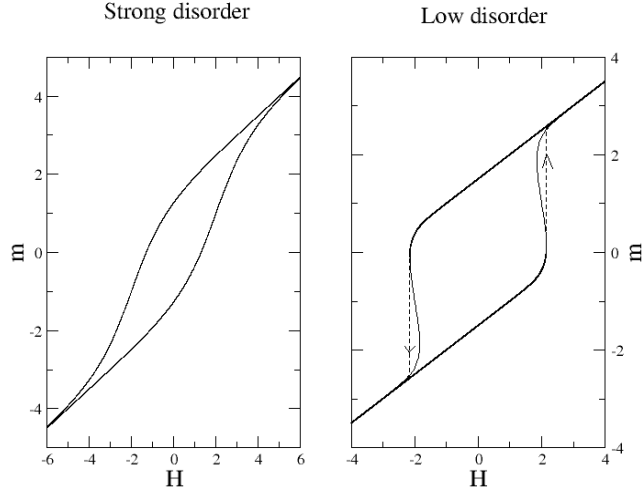

*Figure 1: The soft-spin version of the mean-field RFIM: hysteresis loop (dashed line) and contour $\Sigma_Q(m,H) = 0$ (solid line) for $\Delta = 0.8$. For strong disorder the two curves coincide. For small disorder the contour $\Sigma_Q(m,H) = 0$ is re-entrant and the magnetization curve has a finite jump.*

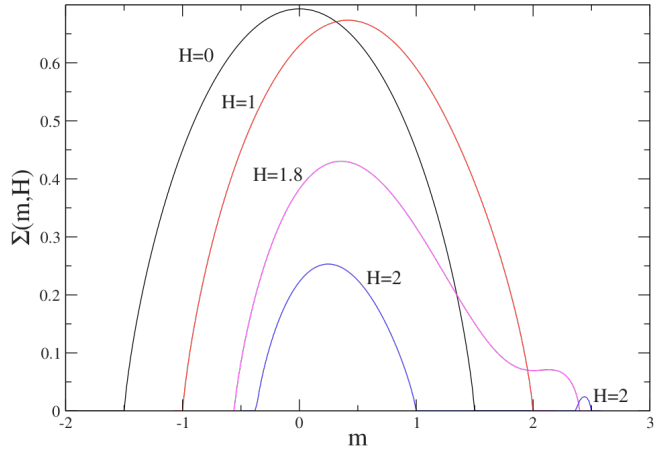

*Figure 2: Complexity vs. magnetization for different values of the magnetic field in the low disorder regime.*

In the small disorder regime the shape of the curve $\Sigma$ vs. $m$ changes drastically with $H$. For $H = 0$, the complexity varies continuously with $m$, reaches a maximum at $m = 0$ (which is thus the most probable magnetization of the metastable states), and vanishes at $m \approx \pm 1.5$, which are exactly the values of the magnetization along the two branches of the hysteresis loop. On the other hand, for a larger field (e.g., $H = 2$), the accessible magnetization domain breaks into two disjoint intervals with no metastable states in between (in the interval $1 \leq m \leq 2.35$ for $H = 2$): this is at the origin of the finite jump in the ascending branch of the hysteresis loop, as can be seen in Fig. 1. One may notice some resemblance of this scenario (a phase transition induced by a disconnected order parameter space) with the ergodicity-breaking scenario observed in systems with long range interactions (see e.g. [14]). However, in the present case, we believe that this is not a consequence of the mean field character of the model and that this scenario is very general.

These results have some interesting consequences. First, on the theoretical side, because they bring up the possibility of studying the hysteresis loop without following the dynamical evolution, which may prove useful to resolve the pending issue of the universality of equilibrium and nonequilibrium disorder-induced phase transitions (there are indeed compelling numerical evidence that the two transitions belong to the same universality class [15] but there is no convincing theoretical proof so far). Secondly, because the distribution of the metastable states in the field-magnetization (or strain-stress) plane also gives a rationale for the influence of the driving mechanism on the non-equilibrium hysteretic response. This is an issue of practical relevance, as we shall now discuss.

2) Influence of the driving mechanism and the effect of long range forces:

A solid bar can be put into tension by specifying either the load that is placed upon it (or hangs from it)–a 'soft' loading device, or its elongation – a 'hard' loading device. In a ferromagnetic material, one usually measures the magnetic flux as a function of the applied magnetic field but one can also make the field slave of the magnetization by using some feed-back mechanism [16]. In adsorption experiments, depending on the size of the gas reservoir connected to the experimental cell, the isotherms could in principle evolve from a 'grand-canonical' to a 'canonical' type [17]. More generally, depending on the system under consideration, one may either control the externally applied field (stress, magnetic field, gas pressure…) or the thermodynamically conjugated variable (strain, magnetization, mass of the adsorbed gas…). At equilibrium the response does not depend on which is the control variable, but what happens far from equilibrium when the response to a smooth external driving is a sequence of avalanches that reflect irreversible transitions between metastable states? What are the differences between the two situations? Since there are very few examples in which the two experimental set-ups have been used with the same disordered sample, the hysteresis loops shown in Fig. 3 are particularly interesting. They were obtained with a $Cu_{68}Zn_{16}Al_{16}$ single crystal under strain-driven and stress-driven conditions [18] and a soft machine was especially designed for this experiment to finely monitor the external force due to a dead load hanging from the sample.

The most striking feature of these curves in that almost the entire strain-driven loop is enclosed within the stress-driven one, showing that the dissipated energy is much larger in the second case. Moreover, the strain-driven curve exhibits a yield point upon loading and a re-entrant behaviour that do not exist with the other device in which there is a macroscopic instability when the martensitic transition starts. Although the microscopic mechanisms at the origin of hysteresis are specific to each particular system, it appears that the same general features are observed in other disordered materials undergoing athermal first-order transition, for instance in magnets [16].

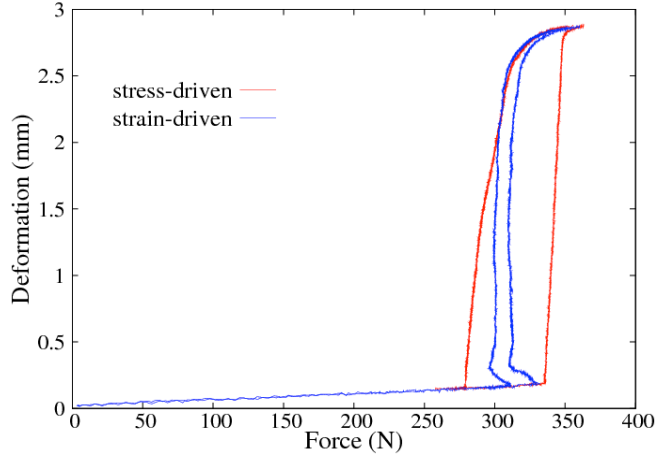

*Figure 3: Stress-strain hysteresis loops in a Cu-Zn-Al single crystal obtained under stress-driving or strain-driving conditions.*

Analysing the soft-spin random-field model is again helpful to reach a global (though admittedly crude) interpretation of the experimental observations (note that Ising spins are inappropriate to study a hard driving situation because the energy wells have no finite curvature, which induces a degenerate and unphysical behaviour [19]). When the magnetization is controlled, the system tries to (partially) minimize its internal energy while satisfying the global constraint $\sum_i s_i = Nm$. It thus visits a sequence of single-spin-flip stable states that is different from the one visited in the field-driven case. In the mean-field model, it can be shown that Eq. (5) is now replaced by

$$s_i - sign(s_i) = m + \frac{h_i}{k} - \frac{1}{N}\sum_j sign(s_j) \ .$$

(6)

The last term in the right-hand side of this equation is an *anti-ferromagnetic* contribution which plays the role of an infinite-range demagnetizing field. Such a field is often introduced to mimic the effects of boundaries or other long-ranged interactions [20]. It changes the system behaviour drastically and leads to self-organize criticality (whereas criticality in the standard field-driven RFIM requires a fine tuning of the disorder). Eq. (6) shows that this is also a natural ingredient of a hard-driving device. A more sophisticated version of this argument can be found in [21] where a disordered spin model is introduced whose critical behaviour changes continuously as one moves from soft to hard driving.

It turns out that the response of the system can be determined exactly when using a very natural relaxation dynamics that states how to go from a metastable state solution of Eq. (6) to the nearest one when $m$ is changed adiabatically. Remarkably, the response is found to *always* coincide with the contour $\Sigma_Q(m,H) = 0$. In other words, the hard-driving device forces the system to follow the boundary of the domain of existence of the metastable states. In the low disorder regime, where the contour $\Sigma_Q(m,H) = 0$ is re-entrant (see Fig. 1), the magnetization-driven hysteresis loop is thus also re-entrant as observed experimentally. These conclusions are in agreement with numerical calculations performed on the metastable RFIM at finite temperature with a local mean-field theory [22]. We believe that these results reflect the general behaviour of a hard driving device.

Finally, we notice that a very recent work [23] shows the equivalence in the continuum limit of the mean-field RFIM with a demagnetizing factor to the celebrated ABBM model [24] that describes the motion of a single domain wall in a random energy landscape. This unifies two rival mean-field theories and shows that it not necessary to assume the existence of an interface from the beginning. In fact, in the hard driving, such an interface is spontaneously created and the system self-organizes at the critical depinning threshold [21].

III WHAT CAN WE LEARN FROM ACOUSTIC EMISSION DETECTION?

One of the motivations of the seminal work of Ref.[1] was its possible applicability to the description of structural transitions in ferroelastic materials, specifically martensitic transitions. The paper also pointed out the parallelism between the Barkhausen noise in ferromagnets and AE signals. This motivation was later partially forgotten because the model was mainly used to interpret various experimental results in magnetic materials.

Acoustic emission has been used for decades to characterize many different processes [25]. From an engineering point of view the technique has been quite successful in monitoring and preventing mechanical failure in solids; it is nowadays the base of many non-destructive testing tools. We shall here focus on the applicability of the technique to the study of structural phases transition in solids. In some aspects this technique plays a role similar to other characterization techniques like calorimetry or resistivity measurements.

The physics behind the source of AE is still far from being fully understood. When a new domain nucleates or an existing interface moves inside the material, an elastic wave is emitted. It propagates through the material and can be detected at the surface by an appropriate transducer. Typically, the observed pulses are ultrasonic, with frequency components in the range 20kHz-2MHz. Within a continuum mechanics description, an AE event can be naively modelled as the sudden creation of a displacement discontinuity [26]. But little is known about the dynamics (acceleration, duration, etc.) of this displacement. A promising recent work [27] may help to clarify this issue.

If the dynamics of this source event were known, the integration of Christoffel equations would allows to predict the AE waves, just like Maxwell Equations are integrated to predict the electric field induced by a sudden magnetization change in the sample. In the magnetic case, the advantage is that one can use detection coils and apply Faraday's law to predict the induced electromagnetic force, thus avoiding an integration that would be otherwise difficult.

Therefore, from the information contained in the detected AE signals, it is very difficult to recover the information about the source. Many of the works that will be mentioned in the following are based on the very simple idea that the maximum amplitude of the detected signals is proportional to the speed of the advancing front [28]. Other fundamental questions remain difficult to answer, as for instance the problem of the spatial localization of the source.

In the study of structural transitions, acoustic emission experiments essentially exploit two basic techniques:

1) Pulse counting technique:

This technique is mostly used to characterize the transition. It consists in counting the number of avalanches (also called events or hits) per unit time (dN/dt) with an amplitude above a certain threshold. When the transition is driven by changing the temperature or another external parameter, this number of events per unit time (frequency) can be converted into the so-called activity A(T) (number of events per degree, or number of events per force interval, etc..) by simply dividing by the driving rate: $A(T) = (\frac{dN}{dT})/(\frac{dT}{dt})$.

With this technique one gets information about the existence of avalanches and the "density" of metastable states along a particular path and in a certain range of the control parameter. A typical result is shown in Figure 4 in the case of the temperature-driven cubic-tetragonal transition in a Fe-Pd alloy [29]. The curves correspond to a polycrystalline sample (top) and to a single crystal (bottom).

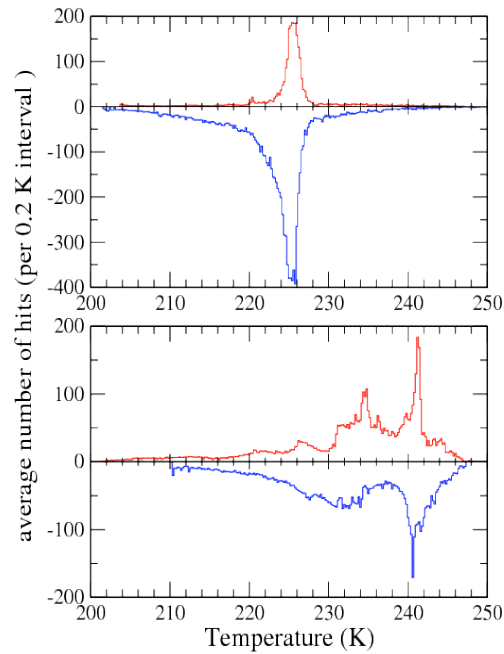

*Figure 4: Acoustic emission activity corresponding to a $Fe_{68.8}Pd_{31.2}$ alloy [29]. The figures correspond to a polycrystalline sample (top) and to a single crystal (bottom). Red lines (positive) correspond to data obtained on heating runs at 1K/min and blue lines (negative) to cooling runs at -1K/min.*

For comparison, Figure 5 shows the calorimetric and susceptibility curves for the same samples.

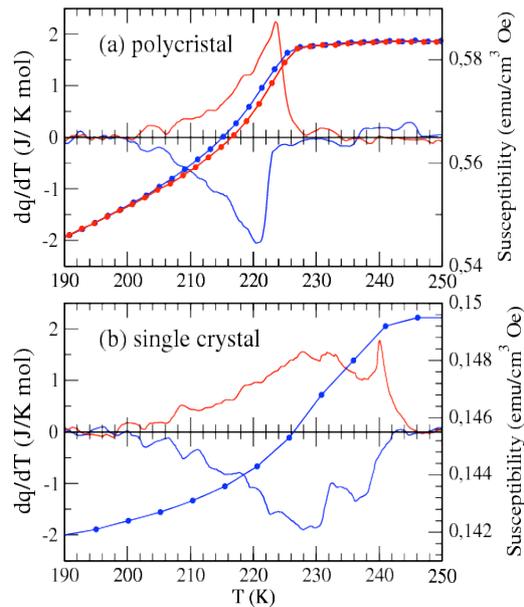

*Figure 5: Calorimetric and susceptibility measurements for the same samples as in Figure 4 [29].*

Although the measurement of the activity may seem to provide little extra information, some important conclusions have been obtained by this technique:

a) <u>Transition temperature</u>: Since AE is much more sensitive than calorimetry, it allows a very accurate measurement of the temperatures at which the transition starts and ends, far beyond the traditional concepts of $M_s$ and $A_f$ temperatures (which correspond to calorimetric estimations of the transformation of 10% and 90% of the sample volume). As an example, Figure 6 shows a magnification of Figure 4 revealing that the starting points of the transition for the polycrystalline sample and the single crystal are the same. It would have been impossible to extract this information from the calorimetric measurements shown in Figure 5.

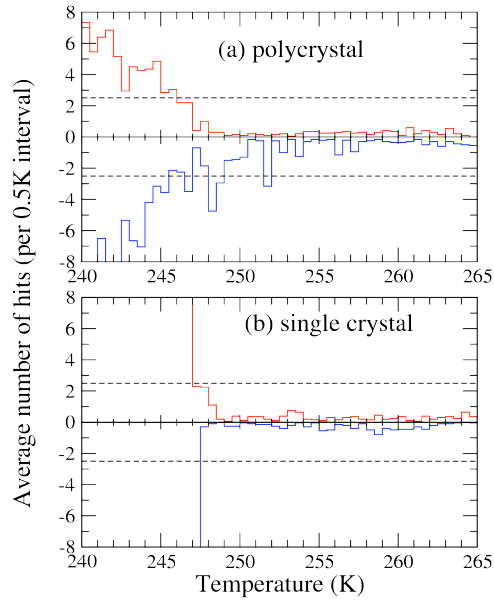

*Figure 6: Magnification of Figure 4 showing that the transformation starts at the same temperature in both samples. The dashed line indicates the noise level [29].*

b) <u>Athermal and adiabatic character of the transition:</u> As was emphasized in section II, in order to exhibit true avalanches a system must behave athermally (thermal fluctuations play no role) and adiabatically (avalanches occur infinitely fast compared to the driving rate). Is it possible to test to what extent these two extreme assumptions are fulfilled in experiments? A first answer comes from the analysis of the dependence of the activity function with the driving rate. As an example, Figure 7 shows the data for a Fe-Pd single crystal recorded at three different rates. Although the overlap is not perfect it is clear that many features of the curves remain unaffected by a change in the driving rate by two orders of magnitude. This overlap (or scaling) is a clear signature that both assumptions (athermal and adiabatic) are satisfied within the range 0.1 K/min – 10K/min.

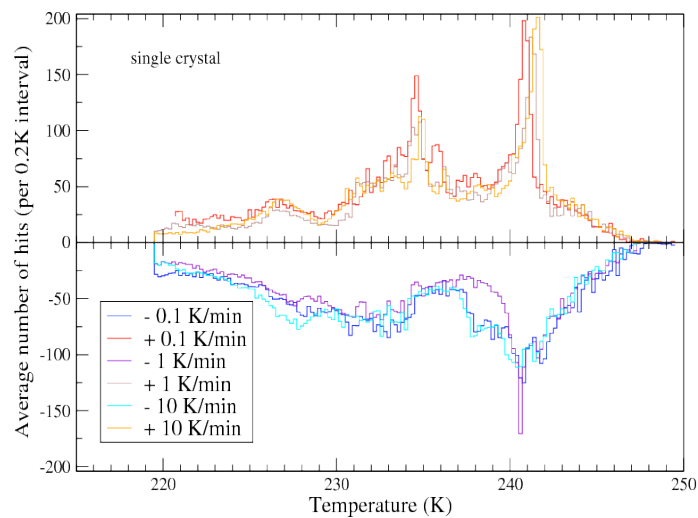

*Figure 7: Acoustic emission activity for different driving rates revealing the athermal and adiabatic character of the structural transition in a $Fe_{68.8}Pd_{31.2}$ single crystal [29].*

However, the scaling is expected to only occur in a certain range of the driving rate. It will not be observed at high driving rates due to the overlap of the avalanches that will necessarily reduce their number (non-adiabatic behaviour). It will also be lost at very slow driving rates due to the occurrence of thermal relaxations (non-athermal behaviour). The activity will then be rate-dependent since the slow driving will increase the probability of thermal relaxation. These upper and lower bounds for the driving rate are often inaccessible experimentally. In some samples, however, it has been possible to observe such changes of behaviour and estimate the degree of "athermaliticity" [30,31].

c) Learning [32]: Two-way shape-memory is one of the most interesting properties of some ferroelastic materials exhibiting avalanche dynamics. This property arises from the interplay between the structural transitions and the reorganization of disorder in the system, and only shows up after a convenient training process. What can be learned about the training process from the AE analysis? Careful measurements in Cu-based shape memory alloys have been performed to investigate the evolution of the acoustic activity. Single crystals are first heat treated in order to "clean" most of the quenched-in disorder (dislocations, vacancies, etc..). The samples are then thermally cycled through the transition by keeping a well-controlled driving rate and fixed minimal and maximal temperatures. As a quantitative measure of the changes occurring from cycle to cycle, the statistical correlation between the activity curves $A(T)$ corresponding to consecutives loops has been calculated. During the initial cycles after the heat treatment, the correlation between consecutive loops is low. But after approximately ten loops, the activity profile tends towards a stable pattern which exhibits a higher correlation between the successive loops. This result shows that the disorder evolves in such a way that the system reaches a final stationary metastable trajectory, which then becomes reproducible.

d) Dependence on the driving mechanism. The AE pulse-counting technique is not restricted to thermally induced transitions. For instance, the technique was used some years ago to study the strain-driven martensitic transition in Ni-Mn-Ga alloys [33]. The theoretical studies of the influence of the driving mechanism presented in section II indicate that significant differences should be observed in the avalanche dynamics when comparing transitions driven by controlling the force/field or the corresponding conjugate displacement. Figure 8 shows the activity profile as a function of the deformation for a $Cu_{68}Zn_{16}Al_{16}$ sample under stress-driven and strain-driven conditions. The data correspond to the same experiment as in Figure 3. The stress-strain curves are also shown for comparison. As can be seen, there is an increase of acoustic activity associated with the strong yield point. In contrast, in the stress-driven case, there is no yield point and the activity is more homogeneous during the transition.

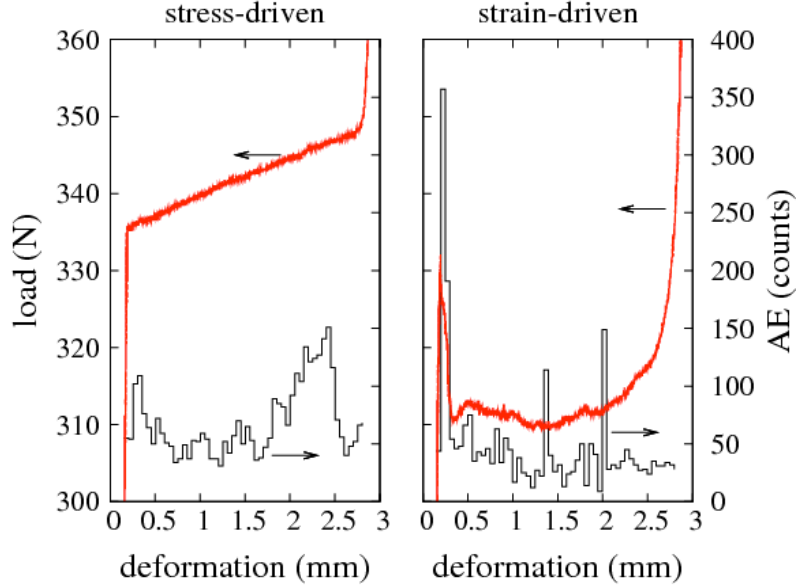

*Figure 8: Acoustic emission activity in a $Cu_{68}Zn_{16}Al_{16}$ sample during stress-driven and strain-driven martensitic transitions [34].*

e) Correlation with calorimetry: An interesting issue which still needs some clarification is the fact that the activity curves correlate very well with the calorimetric curves (see for instance Figures 4 and 5). The main contribution to the calorimetric signal comes from latent heat, and this is naively proportional to the transformed fraction. Therefore the ratio between the activity A(T) = dN/dT and the calorimetric curve dQ/dT should be related to the average volume of the individual avalanches. A similar property has been recently found for the case of stress-induced transitions [35]: the simultaneous measurement of the AE frequency (dN/dt) and the strain changes (dε/dt) reveals a good correlation between both signals. This suggests that it should be possible to define an activity per strain dN/dε, but the low resolution in the stress measurements does not allow to check this point. The rationale behind this interesting hypothesis may be found in some recent results (that will be more commented in the next subsection) that suggest a proportionality relation between the energies of the AE events and the heat released during these individuals events [36].

2) Statistical analysis of single events

The second technique that has been extensively used is based on the detection of a large number of AE signals during the transition and the statistical analysis of their amplitude A, energy E or duration T. These three magnitudes are easily accessible using data acquisition systems. The motivation of this study is to obtain information about the probability densities $p(A)dA$, $p(E)dE$, and $p(T)dT$. If any of these magnitudes A, E or T is related to the avalanche size, and avalanches behave critically (as suggested by theoretical models), one expect that the distributions will exhibit a power-law behaviour

$$p(A)dA \propto A^{-\alpha}dA; \quad p(E)dE \propto E^{-\varepsilon}dE; \quad p(T)dT \propto T^{-\tau}dT$$
(7)

with α, ε and τ being critical exponents.

The main problem for estimating these probability distributions is usually the lack of statistics. To have a good resolution in the highest decades requires an enormous amount of data which is generally not available. Accordingly, in most cases, the statistics is performed by taking into account the signals recorded during the whole transition although the process could be non-homogeneous, as indicated by the activity curves. In fact, some analysis have revealed that there is some change in the histograms when only the initial part of the transition is studied [37,38]. In some cases, in order to gain statistics, one is even forced to average over different cooling or heating runs. This supposes that the system has reached a stationary trajectory after enough cycles.

Once the data are recorded, the sets of amplitudes, duration or energies are analyzed assuming a power-law distribution. In most cases simple histograms show a power-law behaviour with exponents typically ranging from 2 to 4. Figures 9 and 10 show examples of such histograms corresponding to measurements done with a Fe-Pd single crystal.

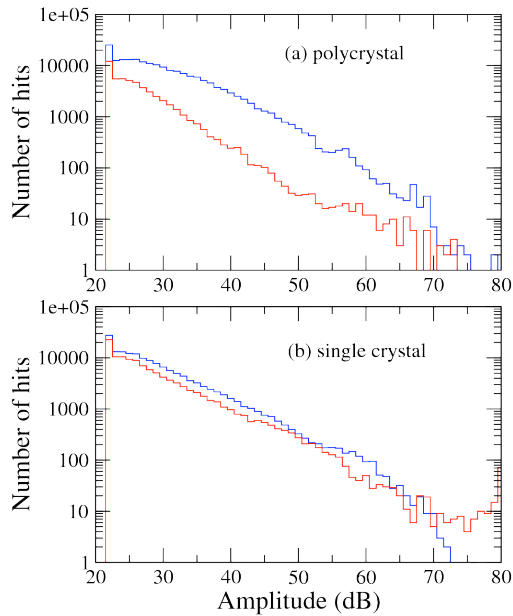

*Figure 9: AE amplitude distribution for two $Fe_{68.8}\,Pd_{31.2}$ samples [29].*

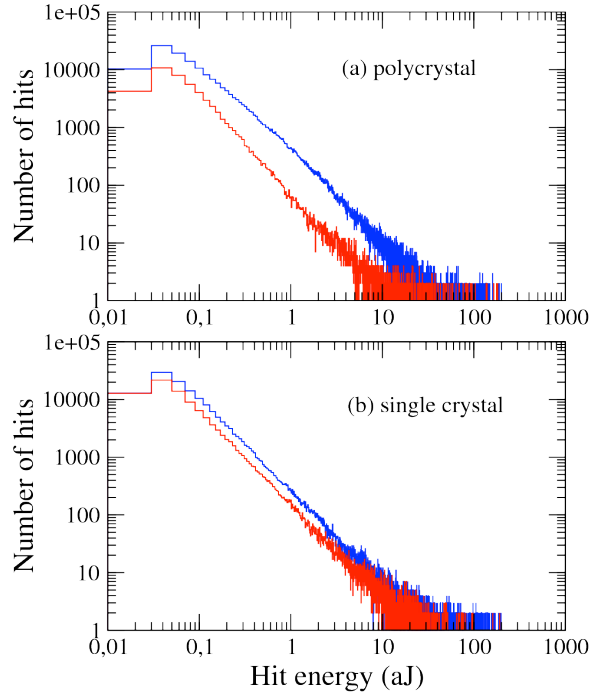

*Figure 10: AE energy distribution for two $Fe_{68.8}Pd_{31.2}$ samples [29].*

To obtain a good numerical estimate of the exponents, the maximum-likelihood fitting methods are used. They provide estimates of the critical exponents and error bars which do not depend on the way histograms are represented. Table 1 presents a compilation of the exponents found in the literature.

*Table 1: Critical exponents obtained from AE experiments in metallic alloys exhibiting martensitic transitions, chronologically ordered. Results in the first line correspond to least-squares estimates (exhibiting large error bars) whereas the others correspond to maximum-likelihood estimates.*

| Material | Transition | Reference | $\alpha$ $p(A) \propto A^{-\alpha}$ | $\tau$ $P(T) \propto T^{-\tau}$ | $\varepsilon$ $P(E) \propto E^{-\varepsilon}$ | x $A \propto T^x$ | y $E \propto A^y$ |
|---|---|---|---|---|---|---|---|
| CuZnAl | Cubic(bcc)-monoclinic(18R) | [37] | 3.6 ±0.8 | 3.5 ±0.8 | | 1.0 ±0.1 | |
| 4 different alloys | Cubic(bcc)-monoclinic(18R) | [39] | 3.1 ±0.2 | | | | |
| 4 different alloys | Cubic(bcc)–orthorhombic(2H) | [39] | 2.4 ±0.2 | | | | |
| CuZnAl | Cubic(bcc)-monoclinic(18R) | [31] | 2.8 ±0.9 | | | | |
| CuAlNi | Cubic(bcc)-orthorhombic(2H) | [31] | 2.48 ±0.7 | | | | |
| CuAlMn | Cubic(bcc)-orthorhombic(2H) | [32] | 2.27 ± 0.03 | | | | |
| FePd | Cubic(fcc)– | [29] | 2.26 | | 1.64 | | 1.97 |
| Single crystal | tetragonal (fct) | | ±0.1 | | ±0.1 | | ±0.4 |
| FePd Polycryst. cooling | Cubic(fcc)–Tetragonal(fct) | [29] | 2.14 ±0.1 | | 1.59 ±0.1 | | 1.92 ±0.4 |
| FePd Polycryst. heating | Cubic(fcc)–Tetragonal(fct) | [29] | 2.95 ±0.1 | | 2.0 ±0.1 | | 1.95 ±0.4 |
| NiMnGa | Premartensitic transition | [40] | 2.44 ± 0.03 | 4.3 ±1.1 | 1.73 ± 0.02 | | |
| NiMnGa | Martensitic Transition | [41] | 2.6 ±0.1 | | 1.75 ±0.1 | | 2 |
| CuZnAl soft driving | Cubic(bcc)-monoclinic(18R) | [34] | 2.95 ± 0.02 | | 2.24 ± 0.02 | | |
| CuZnAl hard driving | Cubic(bcc)-monoclinic(18R) | [34] | 2.67 ± 0.03 | | 1.98 ± 0.03 | | |
| CuZnAl heating | Cubic(bcc)–monoclinic(18R) | [36] | | | 2.15 ± 0.05 | | |
| CuZnAl cooling | Cubic(bcc)–Monoclinic (18R) | [36] | | | 2.05 ± 0.05 | | |

In addition to the exponents, it is also important to study the correlation between the magnitudes measured for each avalanche so to establish whether they are really independent quantities or not. The usual analysis is done by plotting bivariate cloud maps, like the one in Fig.11 representing the energy E vs. the amplitude A of each individual recorded signal. In many cases the maps indicate a clear power-law statistical relation between the measured variables. For instance, Fig. 11 shows evidence that both magnitudes are related, i.e. $E \propto A^{y}$. In this case it is found that y≈2. These statistical dependences may be contrasted with theoretical results that propose a universal shape function for the temporal profile of the avalanches [41,23]. The problem with such comparisons is again the uncertainties in relating the pulse recorded by the transducer to the source. In particular, the pulses duration T (which is determined by an ad-hoc threshold) is probably very sensitive to the transducer response.

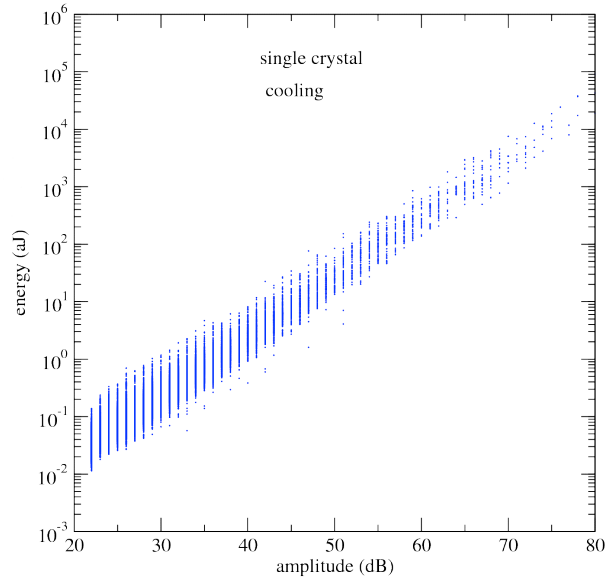

*Figure 11: Energy vs amplitude cloud map of the AE signals recorded in a $Fe_{68.8}$ $Pd_{31.2}$ single crystal [29].*

The main conclusions that have been reached so far from the statistical analysis of individual events can be summarized as follows:

a) Exponent universality classes: For the thermally driven transitions, and provided the driving rates are in the correct regime (athermal behaviour and no avalanche overlap), the values of the exponent $\alpha$ can be grouped in three "universality classes" that depend on the symmetry of the martensitic phase (but not on sample composition): transitions from cubic to monoclinic structure yield $\alpha \approx 2.8$-$3.0$, transitions from cubic to orthorhombic are in general less athermal but yield $\alpha \approx 2.4$-$2.6$, and transitions to a tetragonal structure give an even lower exponent $\alpha \approx 2.3$-$2.4$. A similar conclusion can be obtained for the exponent $\varepsilon$. When the driving rate is too high avalanches may overlap, which decreases the exponent, and when the driving rate is too slow the exponent for the transitions to orthorhombic structure has been found to also decrease. This is because small avalanches become larger due to thermal fluctuations [31].

Some questions remain to be better understood. First, in some cases, deviations beyond the error bars have been found when comparing the forward and reverse transitions. (One should note that the number of recorded signals may be very different depending on the direction of the transition since the activity in the two directions is also very different.) For instance for the $Fe_{68.8}Pd_{31.2}$ polycrystalline sample [29], the exponents corresponding to heating ramps are much higher than expected. This increase is not observed in single crystalline samples. The reason for this deviation could be related to internal strains between grains. Secondly, the application of a magnetic field in the case of NiMnGa samples with a strong magnetoelastic coupling has also been shown to alter the exponents associated with the martensitic transition as well as those associated with the premartensitic transition [40, 41].

Finally, it should be mentioned that a comparison between the avalanche exponents obtained from the analysis of the energy E of AE and calorimetric pulses (at extremely low driving rates) has been performed recently [36]. The values coincide within error bars. This reinforces the idea that the energy is proportional to the heat released in each avalanche.

b) Learning process: The evolution of the exponents with cycling after a heat treatment of the sample has also been studied. The values compiled in Table 1 correspond to samples that have been cycled many times so to reach a stationary AE activity profile. In the initial cycles the fitted exponents show an evolution. In some studies it has even been possible to fit to an exponential correction of the type $p(A)dA \propto A^{-\alpha}e^{-\lambda A}dA$. In this case one finds that $\lambda$ decreases (in absolute value) towards 0 when increasing the number of cycles [32, 39]. Such an evolution of the avalanche distribution towards a stable power-law distribution with cycling has been recently theoretically understood as arising from the interplay between the reversible phase change and the irreversible development of an optimal amount of plastic deformation [43].

c) Influence of the driving mechanism: It has been shown that stress-driven transitions (soft-driving) exhibit exponents comparable to thermally-driven transitions, whereas strain-driving (hard-driving) gives a much higher exponent (due to a smaller proportion of large avalanches). Such an increase of the exponent is in qualitative agreement with the theoretical predictions [21]. It should also be mentioned that the recent measurements with an applied magnetic field [40,41] correspond to thermally induced transitions, but in the near future the same experimental setup may enable to measure AE under magnetic driving.

3) Future trends for the AE technique in the study of structural transitions

There is still much to be learned using the techniques that have just been described. For instance, as suggested by theoretical models, one could analyze AE by performing partial hysteresis loops through the (thermally driven, stress-driven or strain-driven) transition. Comparison with the signal along the main loop could yield some information about the distribution of the metastable states and corroborate the theoretical scenarii described in sections II.1 and II.2. Besides, it is clear that AE waves contain much more information than the one extracted by the above techniques. Many of the methods that are used at large scales (for non-destructive testing or even geological purposes) could be potentially useful for studying structural transitions. In particular, the simultaneous use of several transducers could provide a precise location of the source. This would be a very powerful technique to analyze the dynamics of bulk structural transitions. Location along one dimension has recently been possible in stress and strain-driven samples with a length of 3cm. From this, the energy and amplitude at the source were computed [34]. But locating the source in 2D or 3D is still very difficult at such small scales, given the anisotropic properties of the materials and the complex microstructures generated during the transition. Most probably, numerical simulations in conjunction with experiments will be needed to analyze the information extracted from the AE signals and solve the inverse problem. The simultaneous use of several transducers could also be a useful technique to identify the variant inducing each single AE pulse. This possibility was investigated many years ago [26] but has not been further explored. Finally, as again suggested by theoretical models, it would be interesting to study the statistical distribution of the waiting times between consecutive avalanches. It has been shown recently [44] that they may contain a lot of information that could usefully complement the one extracted from the distribution of amplitudes or energies.

IV CONCLUDING REMARKS

In this review, we have tried to illustrate by some examples the fruitful interaction between theory and experiments over the past few years. Theoretical models, despite their simplicity, have suggested interesting measurements to be made. We also have made some proposals for new experiments. As a final remark, we would like to point out that experiments also suggest that the theoretical description should be improved. In particular, one should study the avalanche properties whose statistical distribution are experimentally accessible and compute the corresponding exponents. So far, focus has been mainly put on the size of the avalanches (volume or number of spins) although this information is still inaccessible in structural transitions. For instance, it would be interesting to study the distribution of the speeds of the advancing interfaces and/or the energy released by each avalanche as this would be closer to the amplitude and energy of the pulses that are actually measured. In this respect, the RFIM is too simple because of the exact symmetry between the parent and product phases which implies the absence of latent heat. Such a symmetry may be valid for some magnetic materials but not for most of other athermal FOPT. Therefore, a more sophisticated model is needed. Although there have been some attempts recently [45,46], there is still much to be done in this direction.

ACKNOWLEDGEMENTS

The authors acknowledge fruitful discussions with Ll.Mañosa, A.Planes, F. J. Pérez-Reche, and G. Tarjus. E.V. acknowledges the hospitality of the Physics Department (University of Warwick) during a sabbatical stay, where part of this review was written. This work has received financial support from Ministerio de Educación (Spain) (PR2009-0016).